\documentclass[twocolumn]{IETEJR}
\usepackage{lineno,hyperref}
\usepackage[linesnumbered,algoruled,boxed,lined]{algorithm2e}
\usepackage{graphicx}
\usepackage{amsmath}
\usepackage{subcaption}

\begin{document}
\title{CARL-DTN: Context Adaptive Reinforcement Learning based Routing Algorithm in Delay Tolerant Network}
\author{ Fuad Yimer Yesuf,and M. Prathap\\
\vspace*{0.05em}\\\small Fuad Yimer. Lecturer, Department of Information Technology, Wolkite University, Ethiopia ,\\\small (corresponding author e-mail: fuad.yimer@wku.edu.et). \\
  \small M. Prathap.  Assistant Professor, Department of Information Technology, \\\small Wollo University, Ethiopia (e-mail: prathapathi@yahoo.co.in).\\  
    }

\twocolumn[
\begin{@twocolumnfalse}
    \maketitle
    \begin{abstract}
   The term Delay/Disruption-Tolerant Networks (DTN) invented to describe and
cover all types of long-delay, disconnected, intermittently connected networks,
where mobility and outages or scheduled contacts may be experienced. This environment is characterized by frequent network partitioning, intermittent connectivity, large or variable delay, asymmetric data rate, and low transmission
reliability. There have been routing protocols developed in DTN. However, those routing algorithms are design based upon specific assumptions. The assumption
makes existing algorithms suitable for specific environment scenarios. Different
routing algorithm uses different relay node selection criteria to select the replication node. Too Frequently forwarding messages can result in excessive packet
loss and large buffer and network overhead. On the other hand, less frequent transmission leads to a lower delivery ratio. In DTN there is a trade-off off between delivery ratio and overhead. In this study, we proposed context adaptive reinforcement learning based routing(CARL-DTN) protocol to determine optimal replicas
of the message based on the real-time density. Our routing protocol jointly uses a real-time physical context, social-tie strength, and real-time message context using fuzzy logic in the routing decision. Multi-hop forwarding probability is also considered for the relay node selection by employing Q-Learning algorithm to estimate the encounter probability between nodes and to learn about nodes
available in the neighbor by discounting reward. The performance of the proposed protocol is evaluated based on various simulation scenarios. The result
shows that the proposed protocol has better performance in terms of message delivery ratio and overhead.
    \end{abstract}
\end{@twocolumnfalse}
]

\section{INTRODUCTION}
Delay/disruption Tolerant Networks (DTN) are a class of wireless network in which at a given instance of time the existence of an end to end path from source to destination is low or it may not exist for the particular time \cite{Fall2003}. DTN provides connectivity in challenging environment, such as wildlife tracking, village communication network, Vehicular Ad hoc Network (VANET), health service for developing regions, satellite communication, social-based mobile network, and Internet of Things (IoT). In such kinds of environment network topology changes frequently, there is no pre-existing infrastructure and end-to-end paths are rarely available due to node mobility behavior and limited capability of mobile nodes (in terms of battery, processing and buffer storage) \cite{Wei2013, Misra2016, Fall2003, Benhamida2017}. 

According to \cite{Zhu2013}, most of the nodes in DTN are mobile, the connectivity of the network is maintained by participant nodes only when they are in the communication range of one another. Due to the node mobility, the network topology is changing frequently, and the assumption of the end-to-end path may not be available temporarily. As a result, routing in DTNs follows a series or relay nodes, this paradigm is known as  Store-Carry-and-Forward paradigm \cite{Zhu2013}. In this paradigm if a node has a message copy but if it is not connected to another node, it stores the message in its buffer until another node comes in to the communication range, once the other node comes in to the communication range it forwards the message to the encountered node with the hope that the encountered node can deliver the message to destination. This process will continue to every encounter node until the message reaches to the destination. When multiple nodes come in the communication range the best relay node selection mechanism will be used according to the routing algorithm design. Different algorithm uses different relay node selection criteria from social network metrics (such as similarity, community, popularity, and betweenness) or pure opportunistic metrics (such as flooding, history of encounter and probabilistic) \cite{Cc2016}.

DTNs communication mainly focuses on sending messages resides in the node buffer to their intended destination by achieving high message delivery and reducing overhead in the network. In this communication environment, there is a thread off between message delivery probability and overhead. Whenever a connection established between two nodes, and if there is a message that needs to be transmitted a decision needs to be made by the routing algorithm runs on the node whether the message should be forwarded or not to the encountered node. Forwarding the message  frequently can result in excessive packet loss and large overhead and buffer utilization. On the other side, less frequent transmission leads to lower delivery ratio \cite{Misra2016}. 

Selection of best relay node which have higher delivery probability towards to destination is one of the main challenges in DTN routing. The probability of successful delivery is dependent on various factors that represent the history and capability of nodes to successfully deliver the message. The node should know the network structure and other nodes available in the network whether there is a sparse or dense connectivity in the area. The network density in a given communication area have higher impact in the message delivery and overhead in the network. Knowing exact density of the node in a given communication area helps to determine the the number of message replicas in a given area to get lower overhead in the network.  

Due to topology change network interruption and limited capability of node affect the amount of packet delivery from source to destination node. To achieve higher message delivery ratio in DTN nodes generate multiple message copy to the encountered node, results additional overhead. As indicated by \cite{Wei2013} in DTN environment there is a trade-off between delivery probability and overhead.

There have been several existing routing protocols proposed in DTN, those routing algorithms are differentiated by queue management, the amount of information available to make forwarding decision,  maximum hop-count a message can have and maximum number of allowed message replicas  in the network. To achieve better performance in terms of delivery probability and overhead, in this study we take into account real-time physical context of node, social-tie strength among nodes, and real-time message context jointly. We use fuzzy logic in the routing decision to prioritize nodes based on their real-time context. Multi-hop forwarding probability is considered using reinforcement learning. 

In this study we design and develop a context adaptive reinforcement learning based routing (CARL-DTN) protocol to determine optimal replica of the message in the network based on the network density. The protocol dynamically detects network environment changes and the current context of nodes to determine a replica of the message, to prioritize the messages and to identify the best relay node in the network.

\section{RELATED WORK}
There are several works which apply machine learning technique to routing in DTN  \cite{Ahmed2010,Rachel2017,Portugal-Poma2014,Sharma2018,ARBR2010,Rolla2013,Wu2018b}.\\
In DTN mobility pattern of nodes shows some level of time periodicity.  The mobility of most real-world DTN follows a repetitive pattern to some extent. The authors \cite{Ahmed2010} propose a routing framework which utilizes the concept of Bayesian classification for determining class membership probabilities by utilizing the network parameter such as spatial and temporal information at the time of packet forwarding. The framework simplifies the integration of various routing attribute and utilizes the repetitive nature of people-centric DTN in making a better routing decision. The framework generally works in two phases, the first phase is the classification phase in this phase the grouping of node based on certain prior information about the network. The second phase packet forwarding can be done using Gradient routing to the neighbor node with the higher relationship with the destination by hoping that the packet will be reached to the destination.

The method explored in \cite{Rachel2017} discuss the concept and architecture of machine learning-based router for interplanetary delay tolerant network. The authors use Reinforcement Learning and Bayesian learning to support the routing decision of contact graph routing (CGR) \cite{Sharma2013}. The authors \cite{Rachel2017} mainly focuses on interplanetary DTN which is a deterministic type of contact (Predicted contact) between each node in the network but it does not consider the non-deterministic type of contact (opportunistic contact).

In \cite{Portugal-Poma2014} authors apply a supervised machine learning algorithm to reduce network overhead by isolating bad relay node for the transmission of message copies. They use decision tree-based classifier to improve routing decision for Epidemic routing \cite{Vahdat2000} by classifying node using attribute vector and delivered classification label. They considered the attribute such as Node ID, region code where the message was received, the message reception time, the lobby index \cite{Korn2009} to measure node density, the time interval between message reception and successful transmission, and distance between where a message received and transmitted to support routing decision. 

In \cite{Sharma2018} proposed a machine learning-based routing protocol for DTN routing called MLProph, they use neural network and decision tree to train based on factors buffer capacity, hop count, node energy, speed, popularity parameter and number of successful deliveries.

 The protocol is evaluated against PRoPHET+ \cite{Huang2010} and the result showed that the proposed protocol superior to PRoPHET+ in terms of delivery probability, overhead ratio, and it shows lower performance in terms of average latency and buffer size due to constraint imposed on the next-hop select process in the MLProph. 

In \cite{ARBR2010} proposed a new collaborative reinforcement learning based routing algorithm in DTN called ARBR. The algorithm assumes nodes cooperate to make a forwarding decision based on contact time statics, node buffer occupancy and congestion sampled during the previous contact between nodes. The node selects the next relay node based on its ability learned from previous contact table exchange. 

In \cite{Rolla2013} the authors proposed a reinforcement learning based routing algorithm in DTN called DTRB. 
The protocol uses multi-agent reinforcement learning learning technique to identify routs in the network to guide message replication process that will produce the best reward.
The protocol assumes the nodes exchange knowledge through regular broadcast control message carries the distance and reward offered for a given message. The protocol calculates the distance table using the gossip-based algorithm. 

In \cite{Wu2018b} proposed a probabilistic vehicular delay-tolerant (VDTN) routing protocol which considers vehicle velocity, social relationship (vehicle mobility and centrality) between vehicle, buffer size and multi-hop forwarding efficiency to make routing decision however it is not enough for Human-centric DTN. The protocol uses vehicle mobility, node centrality and node buffer size for next-hop selection using a fuzzy logic algorithm and the protocol uses Q-Learning \cite{Mitchell1997} to estimate multi-hop encounter probability by discounting reward the with the number of hops from the destination. The protocol assumes each vehicle knows road map information (position and velocity). They use the position and road map information to guide the replication process to make efficient data forwarding. The protocol performs unlimited replication whenever the following conditions satisfies, if candidate vehicle is moving towards the same direction as current vehicle  the data are copied when candidate node shows higher speed than the current node and higher Q-value than the corresponding nodes available in the same road segment and direction, otherwise the data is sent when the candidate node has highest Q-value then the current node. 

\section{PROPOSED ROUTING PROTOCOL}
The architecture of the proposed routing protocol derived from the existing specification of the bundle protocol  RFC5050 \cite{RFC5050} and RFC4838 \cite{RFC4838}. 

\begin{figure}[htb]
		\centering
		\includegraphics[width=0.5\textwidth]{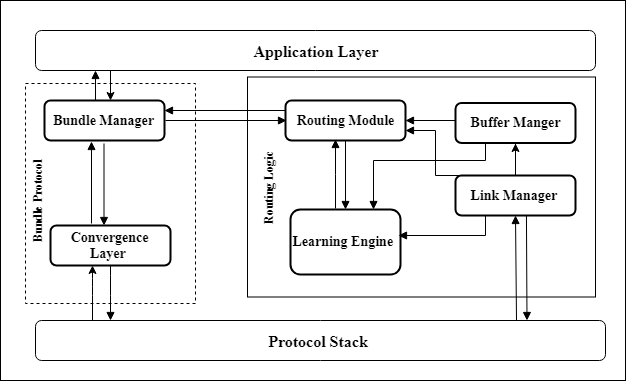}
		\caption[Proposed Protocol Architecture]{Architecture of Proposed Protocol.}
		\label{fig:ProArchitecture}
	\end{figure}

 Figure~\ref{fig:ProArchitecture} shows the general architecture of the proposed routing protocol.

The implementation of bundle protocol has been divided into Bundle Manager and Convergence layer. The bundle layer manager gives the application access to the bundle protocol service and the convergence layer are responsible for interfacing the bundle protocol with lower layers.  The bundle manager is to provide the bundle protocol service to the DTN application running on the node and handle the communication between the application bundle router and convergence layer \cite{RFC5050}. 

\subsection{Link Manger Module}
We assume that each node in the network treated as an equal peer which has sufficient storage and processing capability to make an independent forwarding decision and to learn about network structure and available nodes in the
neighbor. Each node store and maintain a fraction of total network information about the network ( such as Node encounter, remaining buffer size, remaining battery, remaining message TTL and hop-count of message) to identify social characteristics of mobile node, physical context and message context in order to make an independent forwarding decision at each node.

The link manager is used to detect network density in a given communication
range, maintains each node encounter history information, and detect the realtime context of encounter nodes at the time of packet forwarding. When nodes
establish a connection with each other or when the connection is down each
node update their contact information. Finally, the contact information stored
in each node buffer is sent as an input for the learning engine and route module in order to make effective routing decision.

In addition, this module uses the contact history collected using Algorithm~\ref{algo:ensexchange}, then using this information each node identify the social relationship among nodes to improve routing decision.  In this work, we use real-time node popularity and tie strength as a social characteristic to improve forwarding decision and to select the best replication or relay node for the message resides in the buffer.
\subsubsection{Detecting Node Density}
We limit the number of message replicas based on real-time node density in a given area. To achieve the objective the first step we consider is calculating node density to limit the number of message copies accordingly in the current
communication area.

We assume that there are $N$ nodes available in the network. The maximum number of copies for a single message is $L$. In order to achieve a better performance in delivery probability and overhead we should choose a number of message copies $L$ in a given area based on the network density and the capability of the relay nodes to avoid extra overhead. 

As defined by \cite{Liu2014} Encountered Node Set (ENS) is a set of encounter nodes that the current node meets in a period of time $T$. In our design the ENS consists of information about node such as Node Id, meeting time, remain energy, available buffer and connection duration to identify social characteristics of the given node.

Algorithm~\ref{algo:ensexchange} shows the pseudo code for exchanging ENS among nodes in a given communication range to calculate the network density in a given area and social characteristics of the node. 

\begin{algorithm} [!htb]
\DontPrintSemicolon 
\KwIn{Discovery packet generated by node $n1$}
\KwOut{Contact history and real-time context of each nodes available in the communication range}
Node $n1$ waiting for a connection  \;
\For{ Each contact $C$ Available in range  $n1$} {
  \If{$ConnectionIsUP()$} {
   		$n1\gets $Receive each nodes C’s ENS  \;
   		Send $n1$ ENS to each node $C$ available in the range  \;
  
  }
  \If{$ConnectionIsDown() $} {
 		 $n1\gets $updates its ENS 
  }
}
\Return{$ENS$}\;
\caption{Pseudo code for exchange ENS between node}
\label{algo:ensexchange}
\end{algorithm}
As discussed in  \cite{Liu2014} Network density is the ratio between number of encountered nodes and number of nodes in the network. The ENS value is used to estimate the current node density in the current location and used to determine the number of message copies at the time of message creation and forwarding.
E.g., If node $I$ and $J$ encountered with each other, they exchange their ENS using Algorithm~\ref{algo:ensexchange}, the total number of node available in that given area calculated using the formula shown in the Equation~\ref{eqn:totalnoofnode}.

\begin{equation} 
			Total Number of Node= \frac{\left(ENS_{I} \cup ENS_{J}\right)}{N}
			\label{eqn:totalnoofnode}
\end{equation}

\subsubsection{Social Characteristics of Mobile Node}
In this work, we adopt a common centrality measure a   popularity and Tie-strength to measure the activeness of the node for relay node selection.\\

\hspace{2em} \textbf{Popularity }\\
Popularity refers to the number of nods encountered by the node $ i $ in the past $200$ seconds. The popularity of node $i$ calculated using the formula adopted from \cite {Wu2018b} as shown in the Equation~\ref{eqn:popularity}.
\begin{equation}
			Popularity\left(i\right) =\min\left(\frac{NUM_{i}}{NUM_{th} },1\right)
			\label{eqn:popularity}
\end{equation}
where $NUM_{i}$ the number of nodes encountered by the node in the past $200$ seconds. $NUM_{th}$ is the predefined threshold (In this study we use 50 as a default). \\
By tuning parameters, we can set the weight of the centrality factor in making the forwarding decision. $Popularity\left(i\right)$ is then updated for each $NUM_{th}$ period as shown in  Equation~\ref{eqn:updatepopularity}:-

\begin{equation}
\begin{split}
			Popularity_{t}\left(i\right)\leftarrow\left(1-\alpha\right) \times Popularity_{t-1}\left(i\right)+\alpha \times \\ Popularity_{t}\left(i\right)			
			\label{eqn:updatepopularity}
\end{split}
\end{equation}

where t and t-1 shows the current value and previous value respectively.\\

\hspace{2em} \textbf{Tie-Strength }\\
Strong tie between indicates links are more likely to be ready for information flow when compared with weak ties.  For this study we use a combination of tie strength  indication (frequency, closeness and recency) to determine which contact has strongest social relationship to the destination.

Finally, the link manager provides updated node social relationship(popularity and tie strength) information as an input to the routing module and learning engine to select the best relay node in order to increase the delivery probability of a message.

\subsection{Learning Engine}
This study jointly consider the real-time node context (Remaining Energy
and Buffer Status), real-time message status (Message TTL and Message Hop
count) and social centrality metrics (Popularity and Tie strength) in the relay
node selection using fuzzy logic controller and we employ the Q-learning to learn the best multi-hop route.

We assume that every node in the network have limited buffer/storage and processing capability, so each node works as a learning agent and update their own encounter probability for the possible destination when the connection between nodes up or down. When the connection establishes between nodes each node learns the environment by exchanging information with the encountered node. This information helps to select best relay node to forward message resides in the buffer. Learning engine first use the fuzzy logic controller (FLC) to evaluate real time status of node, message status and social relationship of the node then using the Q-learning relay node will be selected for the data forwarding. 

\begin{figure}[htb]
		\centering
		\includegraphics[width=0.5\textwidth]{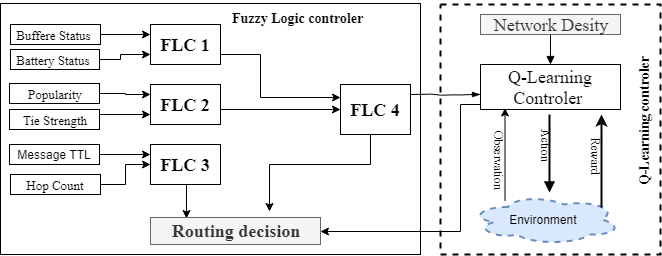}
		\caption[Proposed Learning Engine]{Proposed Learning Engine.}
		\label{fig:learningengine}
	\end{figure}

Learning engine of the proposed protocol is designed using fuzzy inference system and Q-learning as shown in the Figure ~\ref{fig:learningengine}. The proposed FLC designed based upon six input parameters i.e. real-time node context (Remaining Energy and Buffer Status), message status (Message TTL and Message Hop count) and social centrality metrics (Popularity and Tie strength). These six input parameters are applied as an input to three fuzzy inference system labeled as FLC1, FLC2 and FLC3 in the Figure ~\ref{fig:learningengine}.

 The output from first (FLC1) and second (FLC2) fuzzy logic controller become an input for the fourth fuzzy logic controller (FLC4) and gives the final output which is the quality of the node for being a relay node. The FLC 3 (Message context factor) evaluate the real-time status of the message and used in the routing decision to prioritize the messages. 
	
The real-time status of node, message and network is jointly evaluated using fuzzy logic inference system. The Q-Learning algorithm used to learn the best route for the data transmission after computing the fuzzy controller output. The Q-learning uses the fuzzy output value to update the Q-value for each one hop neighbor.

\subsubsection{Fuzzy Logic Controller (Physical and Social context of Node and Message
Context Evaluation)}
\paragraph{\textbf{Node Context Factor (FLC1)}}
The input variable to this fuzzy logic controller are buffer status and battery status, with the linguistic variables \{High, Low, Medium\} for both inputs to represent the node buffer and battery.

Next we use triangular membership function to convert the inputs into fuzzy value. The first fuzzy rule system uses any of nine fuzzy rules shown in the  Table~\ref{tab:ncfrulebase} to handle all possible input scenarios and provide the correct defuzzified output using COG defuzzification method.
\begin{table} [htb]
\caption{Node context factor rule base.}
\label{tab:ncfrulebase}
\centering
\begin{tabular}{c c c}
\hline
\textbf{Buffer Size} & \textbf{Battery Status}  & \textbf{Ability Node}\\
\hline
High & High & Perfect\\
High & Medium & Perfect\\
High & Low & Bad\\
Medium & High & Perfect\\
Medium & Medium & Good\\
Medium & Low & Bad\\
Low & High & Good\\
Low & Medium & Bad\\
Low & Low  & Very Bad\\

\hline
\end{tabular}
\end{table}

The output from this fuzzy logic system is the real-time performance of candidate node for being the relay node or not. The linguistic variable for node real-time performance are defined as \{Perfect, Good, Bad\}. The output value for this fuzzy logic system used as an input for FLC4.

\paragraph{\textbf{Social Importance Factor (FLC2)}}
The input variable to this fuzzy logic system are node  popularity and tie strength. The linguistic variable of node popularity is defined as \{Fast, Medium and Slow\} and similarly the linguistic variable of node tie strength is defined as \{Good, Fair and Poor\}.

Next we use triangular membership function to convert the input into fuzzy value. The fuzzy rule system uses any of nine fuzzy rules as shown in the Table~\ref{tab:ssfrulebase} to provide the correct defuzzified output. 
	\begin{table} [htb]
\caption{Social Similarity rule base.}
\label{tab:ssfrulebase}
\centering
\begin{tabular}{c c c}
\hline
\textbf{Popularity} & \textbf{Tie-strength}  & \textbf{SocialImportance}\\
\hline
Fast & Good & Perfect\\
Fast & Fair & Good \\
Fast & Poor & Good\\ 
Medium & Good & Good \\
Medium & Fair & Good\\
Medium & Poor & Bad\\
Slow & Good & Good \\ 
Slow & Fair & Bad\\
Slow & Poor & Bad\\
\hline
\end{tabular}
\end{table}

The output from this fuzzy logic system is the social importance degree of the node for being relay node with the linguistic variable defined as \{Perfect, Good, Bad\}.

The output value for this fuzzy logic system used as an input for FLC4. Social importance of the node also used in the in the routing decision. 

\paragraph{\textbf{Message Context Factor (FLC3)}}
In this FLC message TTL and Message hop count are used as an input with  linguistic variables \{Large, Medium, Small\} for both inputs.

Next we use triangular membership function to convert the input into fuzzy value. The fuzzy rule system uses any of nine fuzzy rules as shown in the Table~\ref{tab:mcfrulebase} to provide the correct defuzzified output. 

\begin{table} [htb]
\caption{Message context factor rule base.}
\label{tab:mcfrulebase}
\centering
\begin{tabular}{c c c}
\hline
\textbf{Msg TTL} & \textbf{Hop Count}  & \textbf{Msg Priority}\\
\hline
Large & Large & Normal\\
Large & Medium & Normal\\
Large & Small & Low \\
Medium & Large & Normal \\
Medium & Medium & Normal\\
Medium & Small & Low\\
Small & Large & Urgent \\
Small & Medium & Urgent \\
Small & Small & High\\
\hline
\end{tabular}
\end{table}
The output from this fuzzy logic system is the priority of the message to be forward or wait for another relay node with the linguistic variable \{High, Normal, Low\}.

The output of this FLC also used in the routing decision for buffer management and to prioritize the message. Message with low priority transmitted if there are no high priority message in the buffer according to the fuzzy rule. 
 
 \paragraph{\textbf{Transfer Opportunity (FLC4)}}
 The input variable to this fuzzy system are the ability of node to forward a message with a linguistic variable as \{Perfect, Good, Bad\}, and social characteristics of node with linguistic variable defined as \{Perfect, Good, Bad\}. The fuzzy rule system uses any of nine fuzzy rules to handle all possible scenarios and accordingly provide the correct defuzzified output as shown in the Table~\ref{tab:transferopportunity}.
\begin{table} [htb]
\caption{Message context factor rule base.}
\label{tab:transferopportunity}
\centering
\begin{tabular}{c c c}
\hline
\textbf{Node Ability} & \textbf{Social Char}  & \textbf{Transfer Opp.}\\
\hline
Perfect & Perfect & Very High\\
Perfect & Good & Very High \\
Perfect & Bad & Medium\\
Good & Perfect & High \\
Good & Good & High\\
Good & Bad & Low\\
Bad & Perfect & Medium\\
Bad & Good & Low\\
Bad & Bad & Low\\
\hline
\end{tabular}
\end{table} 

 The output of this FLC is the transfer opportunity with linguistic variable defined as \{Very High, High, Medium, Low\}.

The Q-learning controller uses the final defuzzified output value to update the Q-value for each one hop neighbor as well as to evaluate the real-time context of the node and message. 

\subsubsection{Q-Learning Controller (Route Evaluation)}
We use a Q-Learning algorithm to the select best route for the data transmission. Each node learns the environment by exchanging information with the node encountered whenever there is a change in the connection and when message is delivered from one node to its neighbor.

The action at each node to select the best relay node for the data forwarding on the current established connection to relay the messages resides in their buffer.  Therefore, according to the Q-learning algorithm the set of nodes which meets the current agent is the possible action allowed at each agent.  Each node maintains the Q-table as a routing table with the value Q (Destination, Next hop) which tells that predicted value for choosing the current encountered node as the next hop to the destination.  Each node in the network starts with initial Q-table. As discuss in \cite{Andrew2018} there are approaches to initialize Q-value, in this study we initialize the Q-value to zero. The node start without any information about the network then through time the node learns about the neighbors by updating its Q-value.

In our proposed routing protocol, the Q-table is updated when there is a change in connection and when one packet is delivered from one node to neighbor node without sending and receiving RREQ to reduce communication overhead in the network.

 \paragraph{\textbf{Updating and Discounting Reward}  }
Each node maintains the real time context evaluation their transfer opportunity value which is $fuzzy(TransferOpportunity)$. The evaluation value is used for the updating of Q-table. After computing the updating Q-table the forwarder node broadcast a request containing its own delivery probability with the destination and overhearing neighbor respond it back with their own delivery probability. The forwarder node can select the neighbor which has maximum delivery probability with the destination.

We propose three different update strategies for the Q-Table. The first update strategy when the connection established between two nodes the Q-Table is updated using the formula shown in Equation~\ref{eqn:mainQ}. 

	\begin{equation}
	\begin{split}
			Q_{c}\left(d,m\right)\leftarrow\alpha\times \\ \left\{R_{d,m}+\gamma \times Fuzz\left(TOpp\right) Max_{y\epsilon_{N_{m}}} Q_{m}\left(d,m\right) \right\}+ \\ \left(1-\alpha\right)\times Q_{c}\left(d,m\right)
			\label{eqn:mainQ}
	\end{split}
\end{equation}

where $d$ is destination node, $N_{m}$ is a set of nodes which node m met before,$\alpha$ is learning rate 0.3 based on the simulation result, $\gamma$ is discount factor between 0 and 1, $fuzz\left(TOpp\right)$ shows the fuzzy logic evaluation of the real-time node context from the current node to the candidate node, $Max_{y\epsilon_{N_{m}}}Q_{m}\left(d,y\right)$  is a maximal Q-value of $m$ to node $d$, and $R_{\left(d,m\right)}$ is reward $R$ is calculated using Equation~\ref{eqn:rewardR}.

\begin{equation}
R=\left\{\begin{matrix}
1, & if c \in N_{d}\\ 
0, & Otherwise
\end{matrix}\right.
\label{eqn:rewardR}
\end{equation}
The second update strategy when the connection is down between the nodes the corresponding Q-value is reduced with the time elapsed using the formula shown in the Equation~\ref{eqn:aging}.
\begin{equation}
		  Q_{c}\left(d,m\right)\leftarrow Q_{c}\left(d,m\right)\star \beta^{k}
\label{eqn:aging}
\end{equation}

where $ \beta$ is aging constant 
and $k$ is elapsed time since connection lost.

Our learning engine is updated whenever it gets notification from link manager and route module, means when there is a change in the connection an agent updates the corresponding Q-value using the Algorithm~\ref{algo:updateQvalue}.

\begin{algorithm} [htb]
\DontPrintSemicolon 
\KwIn{Connection between nodes}
\KwOut{Updated Q-Table and Reward }
  \If{$ChangedInConnection() $} {
  		\If{$ConIsUP()$} {
  			\For{ Each\ neighbor $y$ in $N_{m}$ } {
  				Get $fuzz(TOpp)$ form Fuzzy controller \;
				Update $R\left(d,m\right)$ using  Equation~\ref{eqn:rewardR} \;
				Update $Q_{c}\left(d,m\right)$ using Equation~\ref{eqn:mainQ} 
  			}
  		}
  		\If{$ConIsDown()$} {
  			Update $R\left(d,m\right)$ using Equation~\ref{eqn:aging}
  		}
  }
\caption{Pseudo code for update Q-value when change in connection occur}
\label{algo:updateQvalue}
\end{algorithm}

The third update strategy after the message is transferred between two nodes the message receiver node updates its Q-table using the message sender node Q-table. E.g., when node $S$ transfer message to node $D$, node $D$ updates its Q-table using node $S$ Q-table and vice versa. If the destination node is known by both nodes the Q-value will be updated in both nodes using the higher Q-value and the next hop will be the node with higher value in both tables. Algorithm~\ref{algo:updateQvalueMsgTransfer}. shows pseudo code for the Q-table update strategy when message is transferred.

\begin{algorithm} [!htb]
\DontPrintSemicolon 
\KwIn{Q-table of both nodes S and node D }
\KwOut{Updated Q-Table }
Initialize $Node S Q-Table $, and $Node D Q-Table$\;
  \While{$messageTransfered between node(S,D)$} {
	\If{$isFinalRecipient()$ OR $isFirstDelivery()$} {
 		  \For{ Each $entry: EncounterNode D.Q_Table.entrySet()$ } {
  				$DTNHost temp = entry.getKey();$ \; 
  				\uIf{$S Qtable\ doesn’t\ contain\ node\ temp$} {
                 	$S.Q_Table.put(temp, D, Node D Qvalue);$   \;
  				}  
			    \Else{
     				\uIf{$ S Qvalue < D Qvalu $} {
                 		$S.Q_Table.put(temp, D, D Q-value)$   \;
  					}  
			   		 \Else{
                    		$ D.Q\_Table.get(temp).updateInfo(S, S,\ Qvalue);$ \;
    				}
    			}
  			} 
  	}
  	Repeat all step for the second node \;
  }
\caption{Pseudo code for update Q-value when message transferred}
\label{algo:updateQvalueMsgTransfer}
\end{algorithm}

\subsection{Routing Module}
The route module provides an interface to bundle manager for receiving and forwarding message, provide persistent storage of bundle with different queue management scheme and perform utility-based replication and queue management.

We employ FLC and Q-Learning in the learning engine to combine different parameters required in the routing decision and relay node selection. We use FLC to evaluate the real time performance of the candidate relay node with the most updated network information to achieve better delivery within highly partitioned environment. In addition, we use the FLC to get the real time message priority based on the hop count and remaining TTL value to prioritize the message using fuzzy rule to avoid packet loss due to lack of priority.

The route module runs at every node meeting, it uses the estimated end to end packet delivery by consulting Q-Table resides in each agent. Each node in the network starts without knowing about other node in the network each agent corresponding Q-Table initializes form lower random value between 0 and 1. Once the packet has been sent to the selected neighbor node, the first node starts updating Q-Table. Each update increases the accuracy of the Q-Table after a certain time. 

Whenever the connection established between two nodes the buffer of one contains a message that needs to be transmitted, a decision needs to be made on whether or not the message should be forwarded to encountered node. The encountered node with maximal Q-value for the destination is selected as next hop. 

\subsubsection{Proposed Copy Control Mechanism}
We use a copy control mechanism to control overhead in the network. We adopt the idea from  spray and focus protocol \cite{Spyropoulos2007}. When the message is generated at the source node an initial value of $L$ number of copies is allowed to the message. Half of the value of $L$ for the message can be forwarded to other node if it satisfies the following condition, if the encountered node is the destination node of a message otherwise real-time message context is considered to forward the message. If the message priority is above normal based on the fuzzy value half of the message copy for a message can be forwarded to the encountered node if the encountered node social importance greater than the current node or the encountered node Q-value is greater than the Q-value of the current node.

When the node has only one message copy it forwards the message to other node if the following conditions satisfies. If the encounter node is the destination of the message otherwise if the message priority is high the message is forwarded to the node with higher Q-value to the destination or encountered node with higher social importance to reduce messages drop due to large hop count and TTL expiration. Otherwise if the message has normal or lower priority the message should be forwarded to node with higher Q-value or node with higher social importance than the current node. Then, it deletes the message from the buffer. A pseudo code of the proposed algorithm shows in  Algorithm~\ref{algo:copycontrol}.

\begin{algorithm} [!htb]
\DontPrintSemicolon 
\KwIn{Connection between node $A$ and $B$ }
\KwOut{Message exchange between node, Update Message property}
Node $A$ and $B$ contact each other\; 
Update contact history between Node $A$ and $B$\;
\For{ Each $M$ in list of message resides in node $A $ } {
  	Compute message priority \;
	Compute Node $A$ and $B$ social value \;
	Compute Node $A$ and $B$ Q-value to message $M$ destination \;
  
  	\uIf{$number of copies of M > 1 $} {
  			\uIf{$M$ destination is node $B$} {
  				$Outgoingmessage.add(M,Connection);$
  			}  
			\uElseIf{$M.priority >= Normal$ }{
	   			\If{$Node\ B\ socialvalue > Node\ A\ social value$ OR $Node\ B\ Qvalue > Node A\ Qvalue$} {
  					$Outgoingmessage.add(M,Connection)$
  				} 
  			}  			
  	}  
	\ElseIf{$number of copies\ of\ M == 1$}{
	   		\uIf{$M$ destination is node $B$} {
  				$Outgoingmessage.add(M,Connection);$
  			}  
			\uElseIf{$M.priority >= Normal  $ }{
	   			\uIf{$Node\ B\ socialvalue > Node\ A\ social\ value$ AND $Node\ B\ Qvalue >node\ A\ Qvalue$} {
  					$Outgoingmessage.add(M,Connection)$
  				} 
  			} 
  			delete from node A buffer after message transferred 
  	}  
		
 }
 
\caption{Proposed algorithm copy control mechanism}
\label{algo:copycontrol}
\end{algorithm}

\subsection{Buffer Manager}
The messages are stored in different intermediate node before arriving at the destination. Each node has a message list table that store the detail of each message. In this work we evaluate the real-time priority of a message using fuzzy logic controller, the real-time priority of a message is updated every time when there is change in connection. The message priority value assists to clear out the buffer in the network. Real-time message priority as described in section FLC3 represented as a linguistic variable \{high, normal, low\} indicates that message with higher priority has large hop count and low TTL value on the other hand message with low priority value indicates that the messages are relatively new message with higher TTL and low hop count. 

The decision to forward or drop a buffered message is taken based on the real-time message priority value. The message in the node buffer is deleted if it is an acknowledgment indicating that the message reached to its destination else the message with highest priority resident in the buffer is removed to make a room for new message. Message with highest priority indicates they have a higher chance to reach the destination because there exists a copy of a message in another intermediate node with higher delivery probability and social importance.

\section{Result and Evaluation}

\subsection{Development and Simulation Tool}
The ONE\cite{Keranen2009} simulator along with   jFuzzyLogic \cite{Jfuzzy2012} has been used to evaluate proposed CARL-DTN protocol.
The ONE is used to develop and evaluate the proposed routing protocol.In addition, We use jFuzzyLogic to design and implement the fuzzy logic part of the proposed routing protocol.

\subsubsection{Simulation Scenario}

TheMap-based movement (MBM), Short-Path MapBased (SPMB), Random-Way Point (RWP) and
Random Walk (RWK) movement models used for
simulation. We have created a simulation scenario
that contains six different groups of nodes, two pedestrians, two car groups, and two bus groups. Among
these groups, cars and bus groups are required to run
on the road. Both groups of nodes use the MapRouteMovement model. Whereas pedestrians use MBM,
SPMB, and RWP movement models alternatively to
check the performance of the proposed protocol in
various movement models.

\subsection{Evaluation and Performance Analysis}
We evaluate the performance of the proposed routing protocols based on various simulation time, buffer size, message TTL and movement models.
\subsubsection{Evaluation Result based on Simulation Time}
	The first simulation performed using various simulation time (5000, 10000, 30000, 43000). 
	
	\begin{figure} [htb]
    \begin{subfigure}[b]{0.25\textwidth}            
            \includegraphics[width=\textwidth]{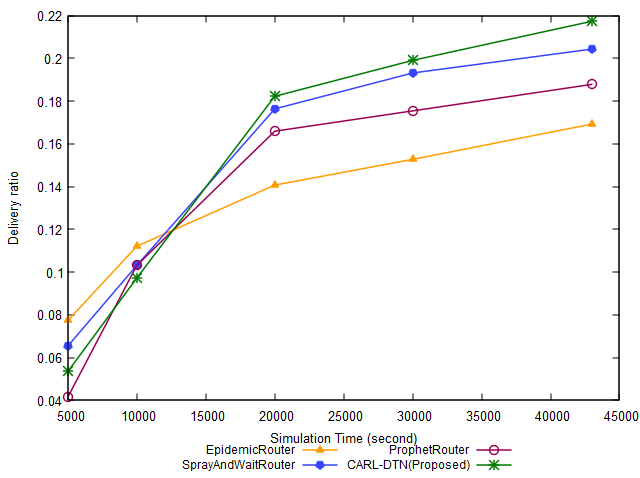}
            \caption{Delivery Probability}
            \label{fig:Dpsimtime}
    \end{subfigure}%
    \begin{subfigure}[b]{0.25\textwidth}
            \centering
            \includegraphics[width=\textwidth]{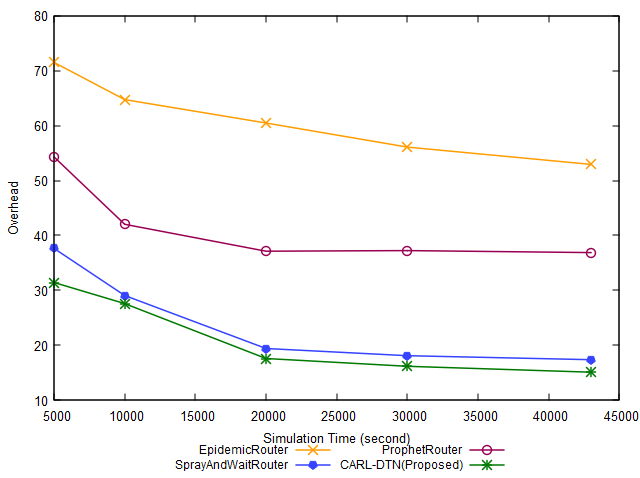}
            \caption{Overhead Ratio}
            \label{fig:ovsimtime}
    \end{subfigure}
    \caption{Effect of simulation time in Delivery and Overhead}\label{fig:SimTime}
\end{figure}

 The results of simulation  are shown in  Figure~\ref{fig:SimTime} two sub figures (A) and (B) shows  comparisons of the Epidemic,PRoPHET, S\&W, and CARL-DTN in terms of message delivery probability and  overhead. 

The result in Figure~\ref{fig:SimTime} shows that the CARL-DTN has improved delivery probability and overhead as compared than Epidemic, PRoPHET,  and S\&W routing protocols. As we have seen in the result Figure~\ref{fig:Dpsimtime} when the simulation time is less than 10000 second Epidemic and S\&W  have relatively a better delivery probability than PRoPHET and CARL-DTN, because Epidemic and S\&W protocols generate multiple copies of the message to the encountered node. On the other hand, when simulation time is greater than 10000 second CARL-DTN has a better delivery probability than the Epidemic, S\&W and PRoPHET. Initially when simulation starts nodes does not have knowledge about the other nodes available in the network through time the node learns about the environment and the algorithm achieves a better delivery probability and overhead.

Epidemic obtains worst performance in-terms of delivery Figure~\ref{fig:Dpsimtime} and overhead Figure~\ref{fig:ovsimtime}, when simulation time increases because of incapability to prioritize the message which is stored in nodes buffer to be dropped or transmit when it is necessary based on real-time message priority. 
 Epidemic routing lets each node keep a copy of each message into its buffer until the message TTL expires and as a result Epidemic introduces the node buffer space to be overflowed, incoming message from new encountered node automatically dropped unless the receiving provides the room for the incoming message by dropping the buffered message from its memory.\\
S\&W distribute all the message copies to encountered relay node quickly in the spray phase and changes in to wait phase to send the received copy to the destination. Both Epidemic and S\&W send  large number of replication than CARL-DTN and PRoPHET algorithm. 

From the result shown in Figure~\ref{fig:SimTime}  PRoPHET shows a better performance than Epidemic, but PRoPHET does not achieve a good performance than the CARL-DTN and S\&W in delivery and overhead. Because of the encounter probability calculation of PRoPHET is based on the history of encounter prediction to select the forwarder node. When the node has no contact with each other encounter probability of the two-node will reduce over time. In addition, the protocol does not consider message priority and node context, whether the encountered node has a performance to deliver the message or not. 

Our Proposed CARL-DTN uses Q-Learning, initially, it takes a little time to learn about network structure it achieves lower performance in delivery when simulation time is less than 10000 seconds but it achieves a better performance than all the three protocols in all simulation time. CARL-DTN jointly considers a real-time node performance, social characteristics  and message priority in the routing decision contributes to achieving a better result in terms  of delivery probability when simulation time greater than 10000 seconds as shown in the Figure~\ref{fig:Dpsimtime}.

The proposed CARL-DTN provides the lowest overhead than all three routing protocols. As shown in Figure~\ref{fig:ovsimtime}, Epidemic has higher overhead than all the routing protocols since it performs unlimited replication, it consumes more network resource, whereas S\& W have a comparable performance with the proposed protocol because it simply performs replication in the initial spray phase of the algorithm after that the node receives a copy waits for the destination this makes the overhead less.

\subsubsection{Evaluation Result for Various Buffer Size}
\begin{figure} [htb]
    \begin{subfigure}[b]{0.25\textwidth}            
            \includegraphics[width=\textwidth]{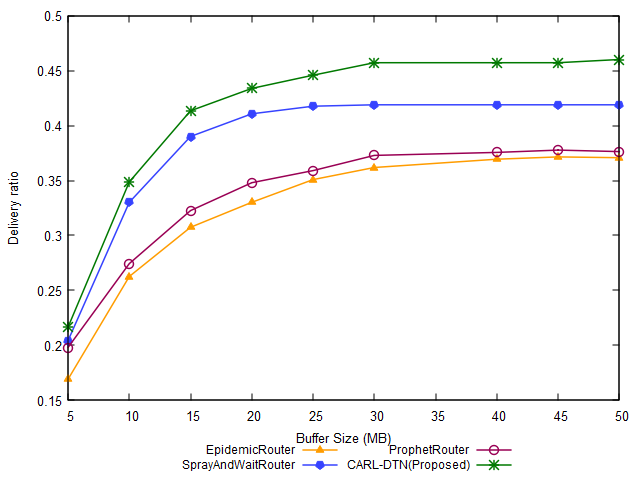}
            \caption{Delivery Probability}
            \label{fig:dpbuffersize}
    \end{subfigure}%
    \begin{subfigure}[b]{0.25\textwidth}
            \centering
            \includegraphics[width=\textwidth]{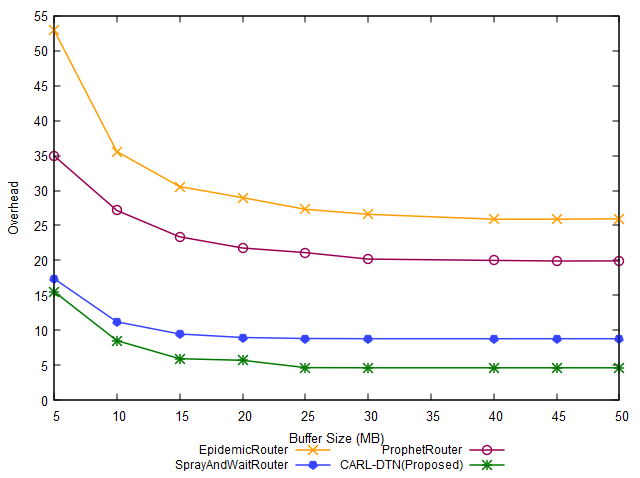}
            \caption{Overhead Ratio}
            \label{fig:ovbuffersize}
    \end{subfigure}
    \caption{Effect of Buffer size in Delivery probability and Overhead}\label{fig:BufferSize}
\end{figure}

Our second simulation performed to evaluate the performance of the protocols using various buffer sizes. Figure~\ref{fig:BufferSize} shows that when the buffer size increases more messages are delivered to the destination at the same time, less overhead is generated.

As shown in  Figure~\ref{fig:BufferSize} the CARL-DTN achieves the best performance in delivery probability and overhead than Epidemic, PRoPHET, and S\&W  especially when the buffer size greater than 5MB as shown in the Figure~\ref{fig:dpbuffersize}. Epidemic and PRoPHET obtain worst performance than CARL-DTN and S\&W in delivery probability and overhead. 
On the other hand, when the buffer size greater than 20MB, both CARL-DTN and S\&W achieves best performance  than other protocol considered in the evaluation and    both protocols are more stable in delivery probability.

Epidemic and PRoPHET protocol allows the node to generate multiple copies of the message in the network which causes high overhead, Epidemic creates multiple messages when the connection established between nodes and if the encountered node does not have the copy of the message. On the other hand, PRoPHET performs unlimited replication when the delivery probability of the encountered node is higher than the current node. The delivery probability in both protocols shows less performance than CARL-DTN and S\&W protocols, due to its lack of potential to prioritize the messages which are stored in nodes buffer to be dropped or transmit when it is necessary. Additionally, both protocol lets each node to keep a copy of each message into its buffer until the messages TTL expires and in turn introduces the nodes buffer space to be overflowed. As a result, the new incoming messages from the new encountered node automatically dropped unless the receiving node provides the room for the incoming messages by dropping the buffered message from its memory.

In reveres, CARL-DTN protocol shows better message delivery probability than the other protocols because of the priority mechanism for the messages stored in the buffer to transmit or wait for the better relay node in terms of delivery probability, social characteristics and physical performance to deliver the message.

As shown in the Figure~\ref{fig:ovbuffersize} CARL-DTN protocol has less overhead than Epidemic, PRoPHET, and S\&W. The CARL-DTN outperforms the Epidemic and PRoPHET in all conditions. As we have seen in Figure~\ref{fig:ovbuffersize}  we can easily observe that when the nodes buffer less than 15MB  our CARL-DTN and S\&W has almost stable overhead. The proposed CARL-DTN protocol uses the real-time message priority mechanism for the messages stored in the node's buffer to be transferred or wait for the best relay node based on the message priority evaluated using a fuzzy logic controller according to the message hop count and TTL. This approach avoids multiple message transmission so that it reduces more overhead for the nodes and also it improves the network performance.

\subsubsection{Performance Evaluation of the Protocols over Various Time-to-Live}
 The third simulation scenario is to evaluate the performance of the proposed CARL-DTN protocol using various message TTL against Epidemic, PRoPHET, and S\&W routing protocol. 
	
	\begin{figure} [htb]
    \begin{subfigure}[b]{0.25\textwidth}            
            \includegraphics[width=\textwidth]{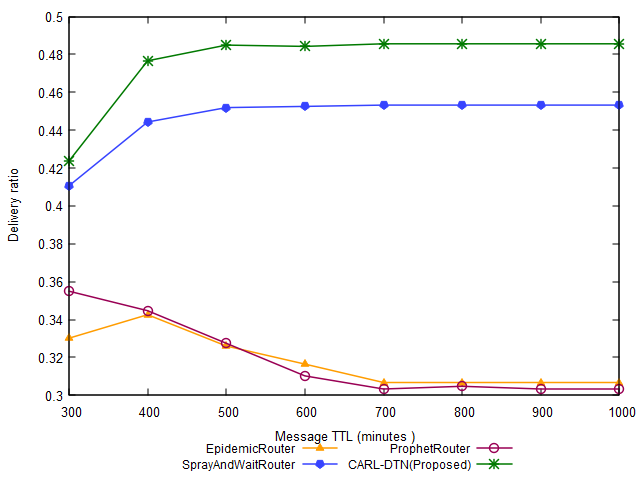}
            \caption{Delivery Probability}
            \label{fig:dpttl}
    \end{subfigure}%
    \begin{subfigure}[b]{0.25\textwidth}
            \centering
            \includegraphics[width=\textwidth]{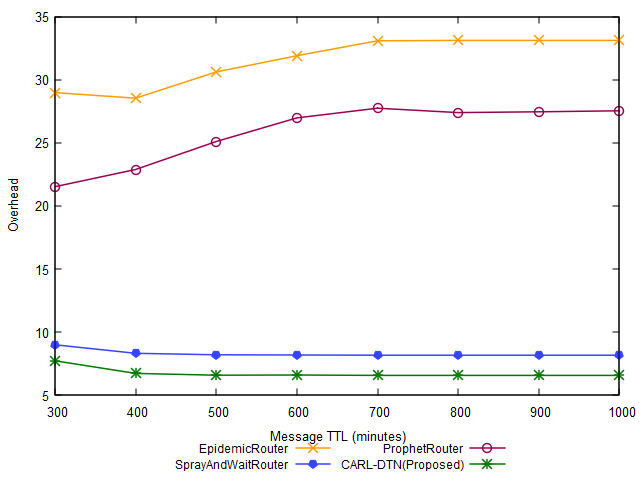}
            \caption{Overhead Ratio}
            \label{fig:ovttl}
    \end{subfigure}
    \caption{Effect of Time-To-Live (TTL) in Delivery probability and Overhead}\label{fig:ttl}
\end{figure}

Figure~\ref{fig:ttl}  shows the performance evaluation of protocols with varying TTL using other simulation parameters specified in Table~\ref{tab:simulationscenario}. As shown in Figure~\ref{fig:dpttl}, when TTL increased the delivery ratio of CARL-DTN and S\&W are increased, On the other hand, the delivery probability  of Epidemic and PRoPHET protocols are decreased. When the TTL increased from 300 min the proposed CARL-DTN outperforms the other routing protocols in delivery probability.

As shown in Figure~\ref{fig:dpttl} CARL-DTN achieves higher delivery ratio than other protocol when message TTL increased from 300 min, because NDNCR has a message priority mechanism designed in the FLC2 to prioritize  message using fuzzy logic based on their  hop count and remaining TTL. In the proposed CARL-DTN messages are forwarded according to hop count and remaining TTL of the message, message with small TTL and large hop count are forwarded first than message with higher TTL remaining and small hop count according to the fuzzy logic rule designed in the FLC2.

Figure~\ref{fig:ovttl} shows that the overhead ratio obtained for different message TTL for Epidemic, PRoPHET, S\&W, and CARL-DTN protocol. As shown in the figure CARL-DTN outperforms Epidemic and PRoPHET algorithms, But the proposed protocol has better overhead than S\&W protocol. Epidemic and PRoPHET have performed unlimited replication until the message reaches the destination. S\&W protocol performs a controlled replication like CARL-DTN, but S\&W does not have a mechanism to select the best replication node. 

To reduce overhead, CARL-DTN removes unlimited replication and each node maintains Q-Table to determine multi-hop delivery probability of a given node in order to select best replication node rather than replicating to every encountered node. This approach extends the life-time of nodes and enhances the probability of successful message delivery by minimizing energy usage required to transmit multiple replications and also by allowing more time for message transmission when node buffer becomes constrained.

\subsection{Evaluation Result for Various Movement Models}
Besides the above three simulation, we also perform a simulation by varying the movement model to check the  performance of the proposed protocol in various movement models.

	\begin{figure} [htb]
    \begin{subfigure}[b]{0.25\textwidth}            
            \includegraphics[width=\textwidth]{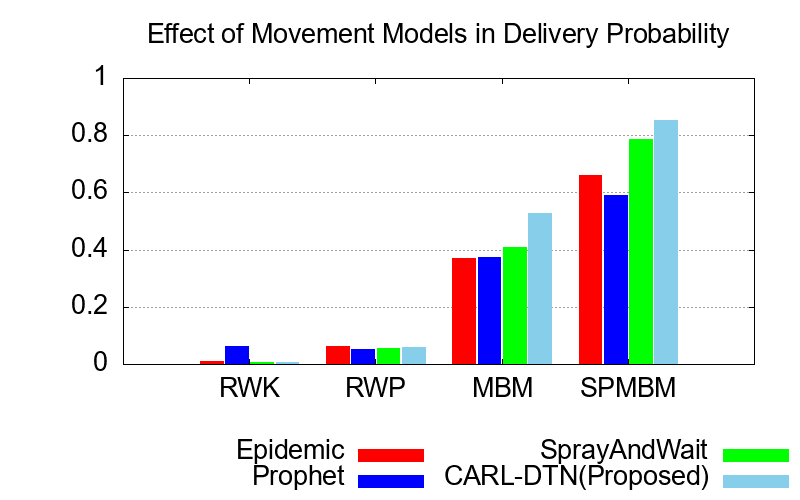}
            \caption{Delivery Probability}
            \label{fig:dpmovtmodel}
    \end{subfigure}%
    \begin{subfigure}[b]{0.25\textwidth}
            \centering
            \includegraphics[width=\textwidth]{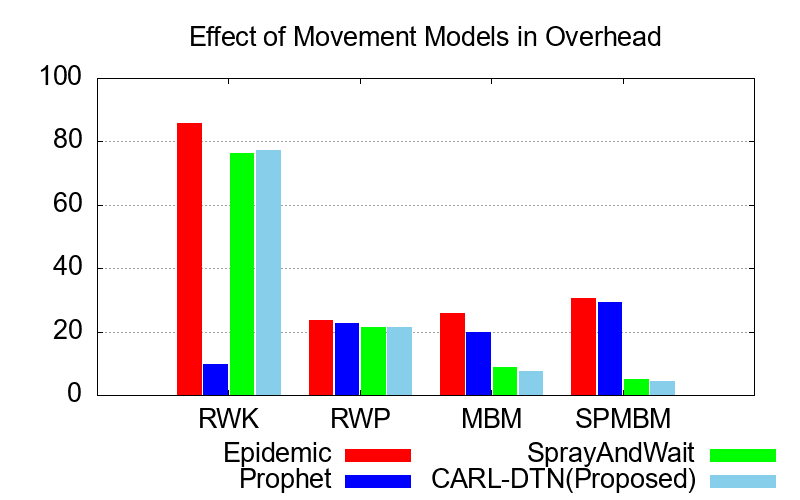}
            \caption{Overhead Ratio}
            \label{fig:ovmovtmodel}
    \end{subfigure}
    \caption{Effect of Movement models in Delivery probability and Overhead}\label{fig:movtmodel}
\end{figure}

As shown in the Figure~\ref{fig:dpmovtmodel} and \ref{fig:ovmovtmodel} the proposed CARL-DTN achieves better performance in MBM and SPMBM movement models than Epidemic and PRoPHET protocol, whereas S\&W has a comparable performance with CARL-DTN protocol in all movement models.

Figure~\ref{fig:movtmodel} depicted that the proposed protocol gets the best result in MBM and SPMBM movement model. However, the proposed protocol achieves a lower result in RWK and RWP movement models than Epidemic protocol. However, CARL-DTN has lower latency and less buffer utilization ratio due to the relay node selection and message priority scheme.

\section{Conclusion and Recommendation} 
\subsection{Conclusions}
In this study, we propose a learning-based probabilistic routing protocol for DTN in order get high delivery probability and lower overhead in the network. We employ the Q-Learning algorithm to estimate the encounter probability between nodes and to learn about nodes available in the neighbor by discounting reward with the time elapsed since the last connection between each node.  The Proposed protocol also uses an adaptive message replication which is able to achieve higher delivery ratio and lower overhead by taking in to account multiple social metrics (popularity, and social ties between node), physical context(remaining battery and available buffer size) and real-time message context (hop count, message TTL ) for estimation of the destination encounter probability and message priority using fuzzy logic controller.

To validate the proposed algorithm, we performed a set of experiments to determine effectiveness of proposed routing algorithm based up on  four simulation scenarios using ONE simulator. Compared to other bench marking routing algorithms from flooding based, controlled flooding, and form history based routing category, our proposed protocol achieves better performance in delivery probability and overhead. 
We observe that from simulation result the proposed protocol shows better performance in terms of delivery probability and overhead in various simulation scenarios. The first simulation scenario is when simulation time grows from 2000 seconds to 45000 seconds. The second simulation scenario when we use various buffer sizes such as 5MB up to 50MB. The third simulation scenario when message TTL varies from 300 minutes to 1000 minutes. Final simulation scenario when we are using various movement models such as MBM and SPMBM. So, our proposed protocol shows better performance in terms of the aforementioned simulation scenario and performance metrics.

This study shows the strength of hybridization, i.e given multiple routing scheme, each suited for a various network scenario of the DTNs design space, it makes sense to capitalize on each protocols strengths by combining them into a single routing protocol which is a better solution for scalable routing.

%
%
%

%
\bibliographystyle{ieeetran}
\bibliography{library}

\end{document}